# Disk galaxy rotation curves and dark matter distribution

by
**Dilip G. Banhatti**
School of Physics, Madurai-Kamaraj University, Madurai 625021, India
(Currently visiting Solid State Theory group at University of Münster, Germany)
[E-mail : dilip.g.banhatti@gmail.com, banhatti@uni-muenster.de]

**Prelude.** Initially called "missing mass" in the 1930s when Fritz Zwicky thought there was insufficient mass to provide the gravity needed in galaxy clusters to balance the measured speeds, and renamed "missing light" four decades later, when it was realized that the clusters are relaxed structures most likely in or near virial equilibrium, so that it was light that was not observed while the gravity was felt, the presence of "dark matter" was soon inferred on all scales from individual galaxies to groups to clusters. In the solar system, Neptune was found at the spot predicted from anomalous motion of Uranus, while the anomalous part of the precession of Mercury did not lead to Vulcan, rather it was explained by general relativity, Einstein's improvement over Newton of the description of gravity. Rotation about the centre of a spiral galaxy, in analogy with planets moving around the Sun, has been used to get the distribution of mass inside the orbits of test particles. Assuming that the sums are done right, the variation of circular speed with galactocentric distance, called the "rotation curve", gives the distribution of mass to provide the gravity needed to hold the test particles in orbits. On page 986, Banhatti presents the current status of these observations and calculations for disk galaxy rotation curves, and the inferred dark matter, for the Milky Way as well as other spiral galaxies.

**Abstract.** After explaining the motivation for this article, I briefly recapitulate the methods used to determine, somewhat coarsely, the rotation curves of our Milky Way Galaxy and other spiral galaxies, especially in their outer parts, and the results of applying these methods. Recent observations and models of the very inner central parts of galaxian rotation curves are only briefly described. I then present the essential Newtonian theory of (disk) galaxy rotation curves. The next two sections present two numerical simulation schemes and brief results. Application of modified Newtonian dynamics to the outer parts of disk galaxies is then described. Finally, attempts to apply Einsteinian general relativity to the dynamics are summarized. The article ends with a summary and prospects for further work in this area.

**Motivation.** Extensive radio observations determined the detailed rotation curve of our Milky Way Galaxy as well as other (spiral) disk galaxies to be flat, much beyond their extent as seen in the optical band. Assuming a balance between the gravitational and centrifugal forces within Newtonian mechanics, the orbital speed V is expected to fall with the galactocentric distance r as $V^2 = GM/r$, beyond the physical extent of the galaxy of mass M, G being the gravitational constant. The run of V against r, for distances less than the physical extent, then leads to the distribution M(r) of mass within radius r. The observation V ≈ constant for large enough r, up to the largest r, up to 100 kpc, thus shows that there is substantial amount of matter beyond even this largest distance. A spherically symmetric matter density $\rho(r) \propto 1/r^2$, characteristic of an isothermal ideal gas sphere, naturally leads to V = constant. Since this matter does not emit radiation, it is called dark matter. In general, the existence of dark matter is, by astrophysical definition, inferred solely from its gravitational effects. (Astro)particle physicists[1] hope to change this by directly detecting dark matter particles. Considering the evidence for different types of dark matter on scales from our solar system out to the observable universe[2], the detailed structure on the sky of cosmic microwave background radiation[3], simulations for large-scale structure formation[4], and big bang nucleosynthesis calculations[5], such gravitationally & otherwise normally interacting dark matter may be made of very light normal particles like (massive) neutrinos which are relativistic (& hence called hot dark matter) or much heavier (about a $GeV/c^2$ or more) exotic nonrelativistic particles (hence called cold dark matter)[6]. By 'exotic' here I mean <u>not</u> any of the particle zoo of the standard model of particle physics[6]. For an exposition regarding our current picture of the composition of the universe and other details, see, e.g., Grupen[1] & references therein. Here, I confine attention to the scale of individual galaxies.

One often assumes an isothermal dark matter halo, although $\rho(r) \propto 1/r^2$ is only one of many density profiles leading to V = constant, the others being disk-like. For disk-like mass distributions, in contrast to spherically symmetric ones, the circular speed at a given r is determined by matter distributed from 0 to r and also beyond r, as can be easily seen by applying Gauss' integral theorem (or law) to appropriately shaped closed volumes. Some textbooks make the error of integrating only up to r, leading to wrong results, as pointed out by Méra et al (1996/7)[7].

---

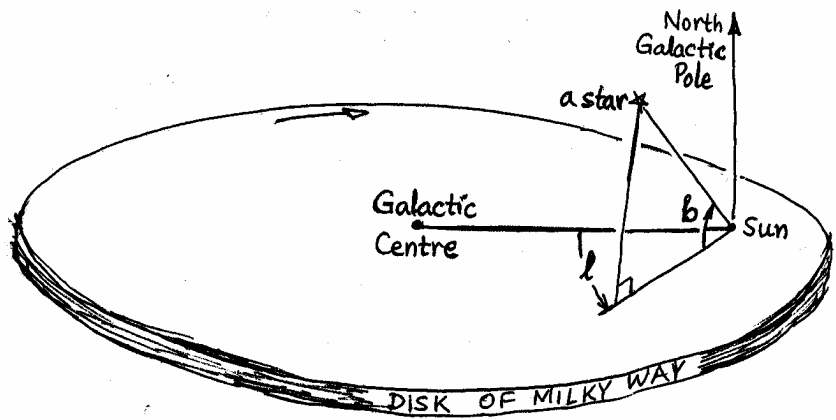

**Figure 1.** A schematic picture of Sun's location in Milky Way Galaxy, illustrating Galactic coordinates b (latitude) and ℓ (longitude). Rotation is indicated by an arrow.

Recently, de Boer et al[8] reanalyzed the public domain data from the 0.1 to 10 GeV all-sky γ-ray survey, which was done by the satellite-borne telescope EGRET. They found that excess diffuse γ rays of the same spectrum are observed in all the sky directions, and that the spectral shape can be interpreted as annihilation of (dark matter) particles (and antiparticles) into intermediaries like $\pi^0$ mesons and then to γ-ray photons, implying a mass of 60 to 70 MeV/$c^2$ for the annihilating (dark matter weakly interacting massive) particle (or antiparticle). From the intensity variation of the diffuse γ-ray excess with respect to Galactic longitude and latitude (Figure 1), and assuming a spherically symmetric component in the mass distribution, they derived an almost spherical isothermal profile plus substructure in the Galactic Plane in the form of two toroidal rings at 4.2 kpc and 14 kpc from the Galactic Centre. (The absolute normalization of the dark matter profile is tied to the local rotation speed 220 km/s at 8.3 kpc, Sun's distance from the Galactic Centre. This may need renormalization for consistency with better and more recent determinations, e.g., like those of Xu et al[9] and Hachisuka et al[10].) The two rings produce, within Newtonian dynamics, the observed bumps in the detailed shape of the Galactic rotation curve. These rings are actually broken segments disposed around the Galactic Centre in ring-like structures, as also seen for OB stellar associations via their distribution and kinematics[11]. In general, nonaxisymmetric structures like spiral arms and bars should also be taken into account[12]. Dekel[13] cautions to choose carefully test particles to measure rotation curves in general, giving an example where low stellar speeds turn out to be a red herring, detailed disks' merger model for the elliptical galaxies in question showing orbits consistent with dark matter halo *and* low stellar speeds. Aharonian et al[14] describe the discovery of TeV γ rays from the Galactic Centre Ridge. Ando[15] treats cosmic γ-rays as being from dark matter annihilation. In this article, I restrict myself to outer parts of galaxies, i.e., to sufficiently large r, where the rotation curve has stabilized to a flat shape on average.

-------------------------------------------------------------

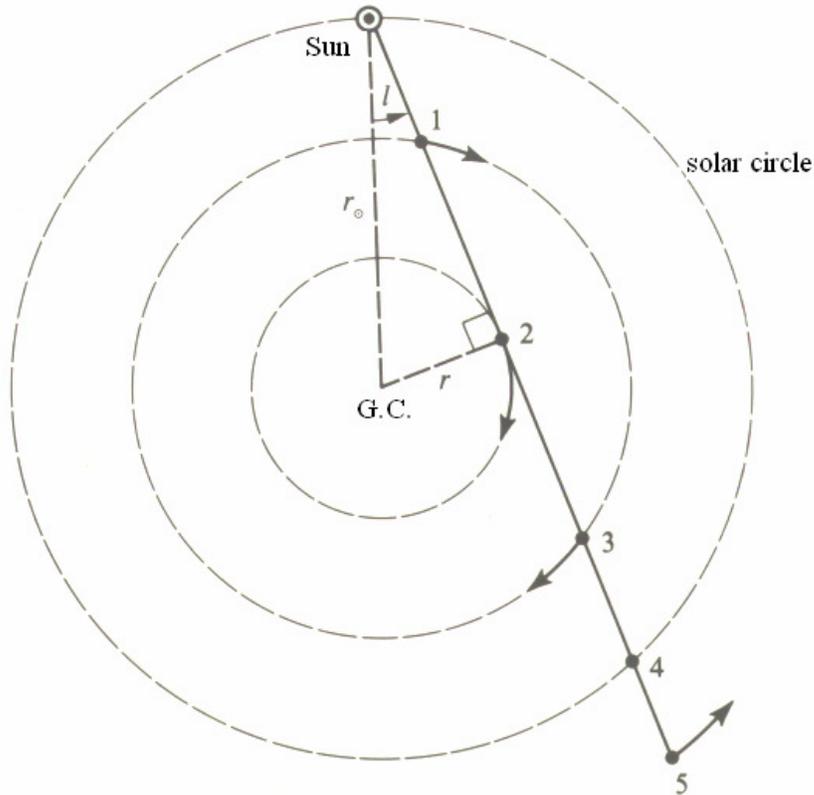

**Figure 2a.** Different types of clouds in a given direction. See Fig. 2b for the line profile.

---

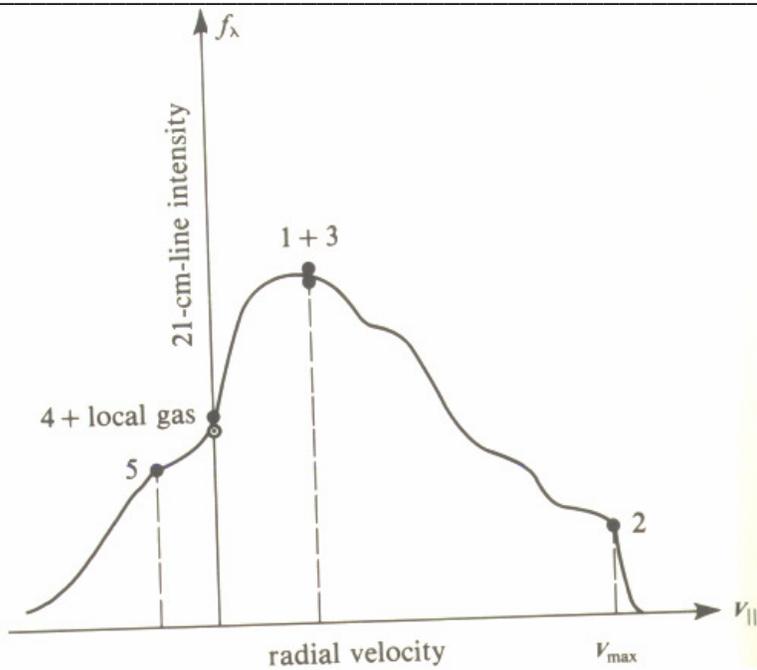

**Figure 2b.** Neutral hydrogen line profile showing typical regions as numbered in Fig. 2a.

---

**Measurements in the Milky Way Galaxy**[16,17]. In the kinematic method of determining distances in the Galactic Plane in a given direction (Figure 2), the radial speeds of neutral hydrogen (i.e., H I ≡ H$^0$) clouds are measured by the Doppler shifts of the $\lambda$21 cm line corresponding to the electronic spin-flip transition of the H atom. The highest radial speed in the line profile in Galactic longitude direction $\ell$ gives the circular rotation speed V at distance r = $r_{Sun}\sin\ell$ from the Galactic Centre. The distance ambiguity for clouds of types 1 and 3 is removed (or resolved) by measuring their extent in Galactic latitude (see Figure 3). Another way is to measure hydrogen absorption at radio frequencies, since distant sources show wider velocity range[18]. Such measurements give the Milky Way

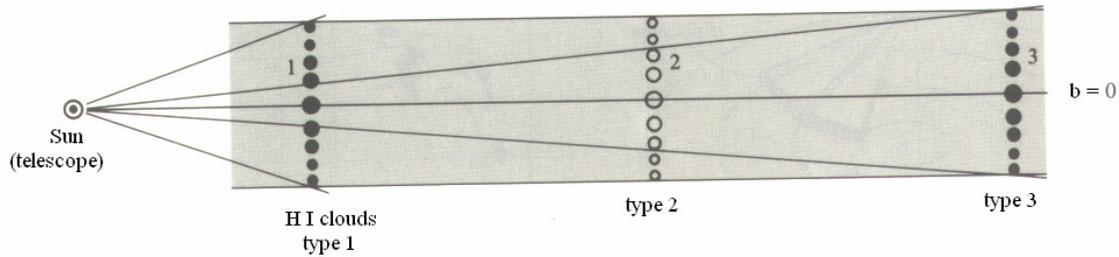

**Figure 3.** Resolving distance ambiguity by comparing widths in Galactic latitude.

rotation curve in its outer parts, up to about the Solar Circle, as shown in Figure 4. Relative to the Sun (i.e., a frame of reference rotating with angular speed $\Omega(r_{Sun})$), the circular speed of a gas cloud at radius r is $r[\Omega(r) - \Omega(r_{Sun})]$ (Figure 5). The line of sight speed $V_\parallel$ at a given Galactic latitude $\ell$ is

$V_\parallel = r[\Omega(r) - \Omega(r_{Sun})]\sin(\theta + \ell) = r_{Sun}[\Omega(r) - \Omega(r_{Sun})]\sin\ell.$

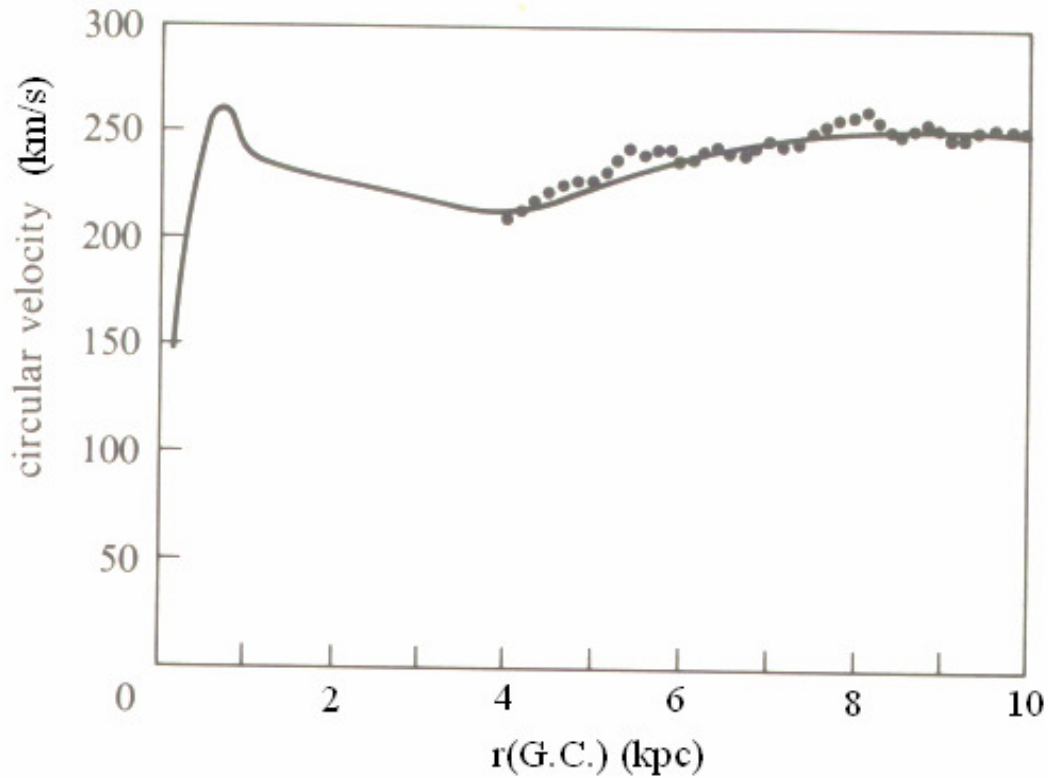

**Figure 4.** Milky Way rotation curve in outer parts

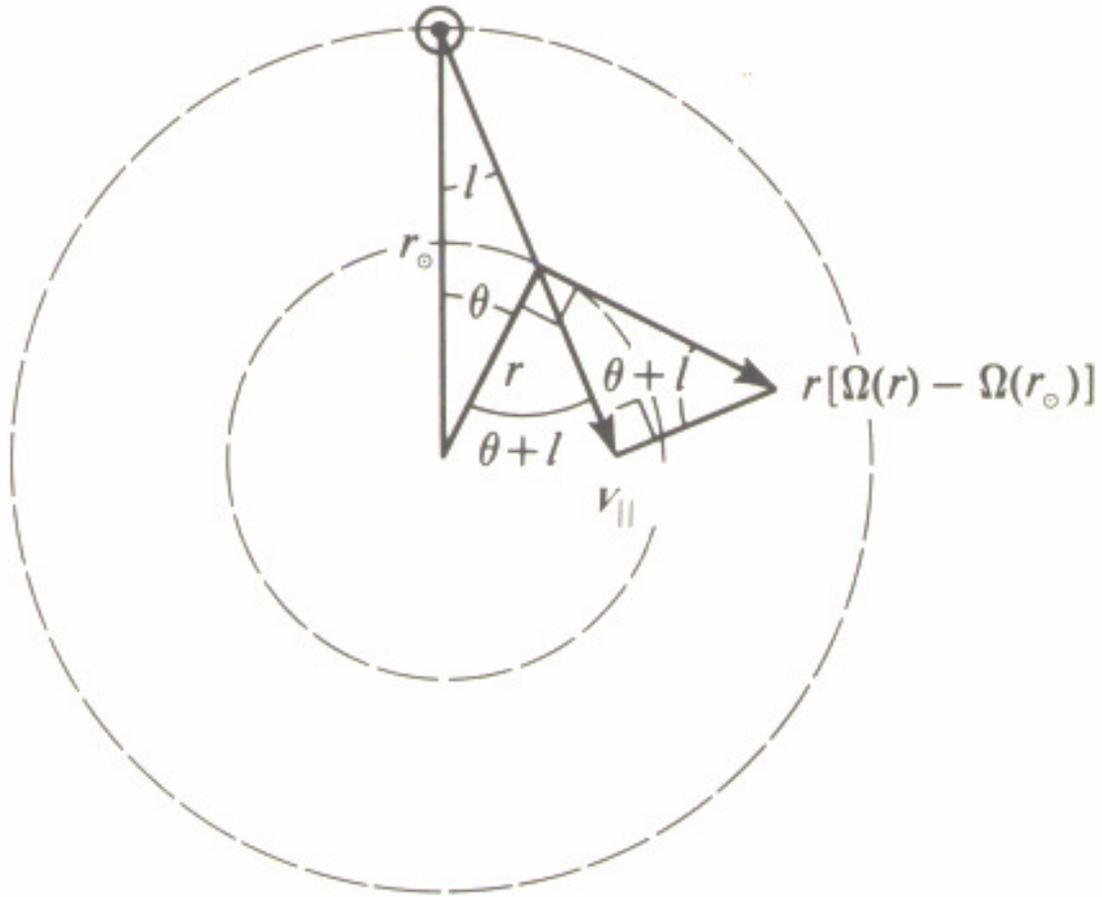

**Figure 5.** Geometry to calculate kinematic distances in the Galactic Plane.

Thus $V(r) = r\Omega(r) = V_{max} + r_{Sun}\Omega(r_{Sun})\sin\ell$, where $V_{max}$ is the maximum value of $V_{\parallel}$ from the line profile. Beyond Solar Circle $r = r_{Sun}$, giant H II ≡ $H^+$ complexes are used in place of H I (≡ $H^0$) regions. The tracers for these complexes are the CO line (at $\lambda \approx 1$ to 3 mm) and H109α (at λ6 cm), among others. Figure 6 shows the Milky Way rotation curve beyond the Solar Circle, with the "expected" curves for uniform (i.e., rigid-body) rotation in the inner part and Keplerian behaviour in the outer part. Coarse observations, when interpolated across the nucleus, correspond roughly to uniform or rigid-body rotation in the inner part, although recently discovered finer scale structure there has other implications[19]. Briefly, molecular and maser line spectroscopy of massive inner nuclear disks has shown the presence of order of magnitude larger Keplerian speeds than the outer 'flat' value, indicating a possible supermassive nuclear black hole. Bar kinematics has been inferred from detailed observations. Even counter-rotating nuclear disks, possibly resulting from mergers, are present. Perhaps the nuclear black hole has 'sucked in' these peculiarities of structure & kinematics to leave more orderly (flat) rotational kinematics in the outer parts of (spiral) galaxies. The reader can refer to the literature for recent observational details[20]. In the outer part, the rotation speed is clearly super-

Keplerian, on average flat, with some modulations, in this plot (Figure 6) from Shu[16]. Binney & Tremaine[17] summarize the detailed disk surface mass density model of Caldwell & Ostriker[21] which is essentially a fit with 13 observationally determined parameters:

$\sigma(R) = \sigma_0[\exp(-R/R_1) - \exp(-R/R_2)]$. The central model density is zero, which cannot accommodate a central mass concentration, for which there is ample independent evidence[19]. However, the model has toroidal rings at $R_1$ and $R_2$ for which there is separate observational evidence[8,11] as mentioned earlier. So a superposition of such a model with another component having centrally concentrated mass should fit the observed rotation curve for inner as well as outer parts of the Milky Way Galaxy. For other detailed models of (especially the outer parts of) the Milky Way rotation curve, see Cowsik et al[22] and Dehnen & Binney[23].

---------------------------------------------------------

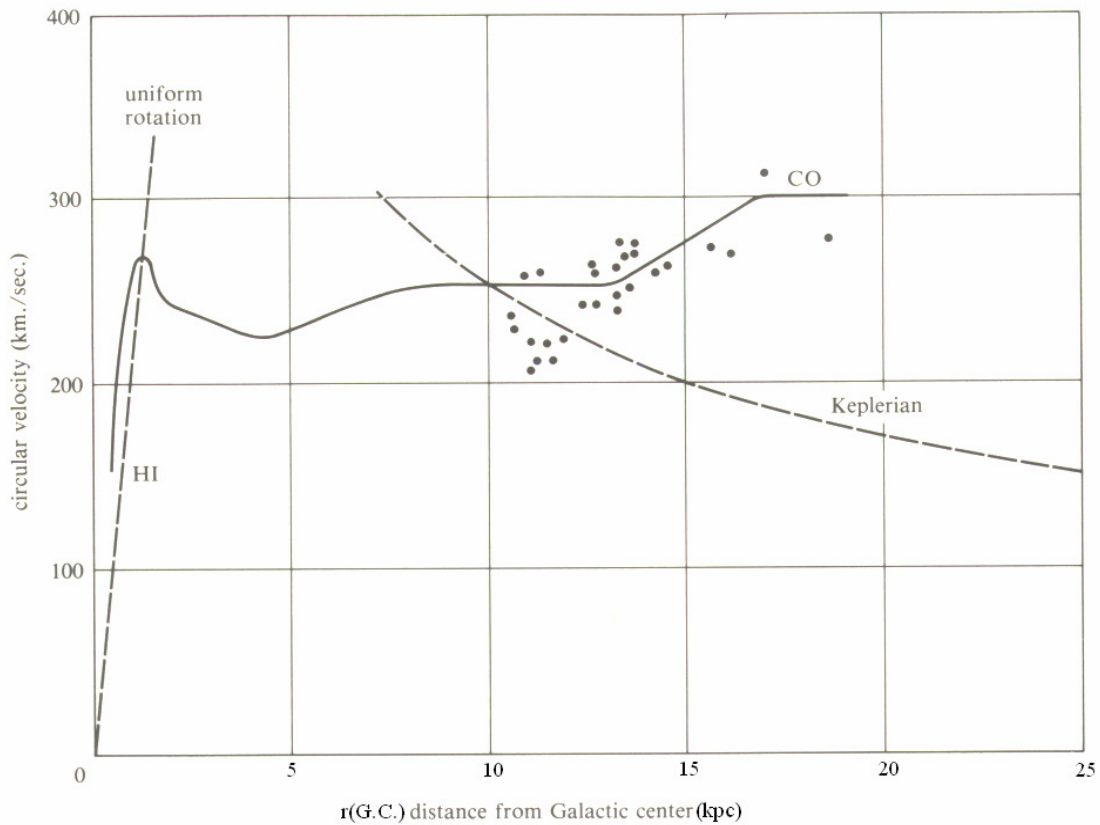

**Figure 6.** Milky Way rotation curve beyond the Solar Circle. Uniform rotation means rigid-body rotation $v = \omega \cdot r$, with uniform angular speed $\omega$, independent of r.

**Rotation curves of other galaxies.** Assuming Newtonian gravitation to predominantly determine the dynamics, the mass M may be estimated from orbital speed V in ellipticals as well as spirals. The mass M interior to r is roughly $M(r) = rV^2/G$, with r the deprojected distance and V the (spread in) random speeds for ellipticals, while in spirals it refers to the circular speed about the galaxian centre. For a disk galaxy it is more meaningful to take M(R) as the mass within a *cylinder* of radius R, while for a spheroidal galaxy M(r) is more conveniently the mass within a *sphere* of radius r. Observationally, measurements are possible only along the total line of sight through a galaxy (Figure 7). Thus the cylindrical radius is more pertinent for observations. However, the practical difference between the two is not large for actual galaxies, as is seen from the following example. Take $\rho(r) = C/r^2$. Then the surface mass density is

$\sigma(R) = \int_{all\ z} dz\ \rho(r)$, where $r^2 = R^2 + z^2$ (Figure 7)
$\quad\quad = \pi C/R$.

---

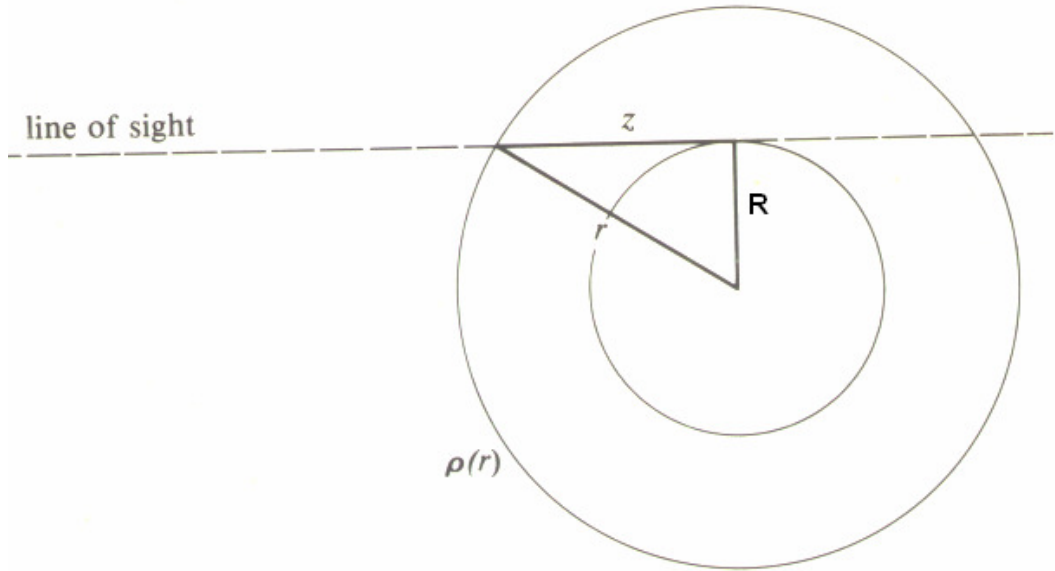

**Figure 7.** Line of sight through a galaxy, showing relation between spherical radius r and cylindrical radius R.

Hence the masses within a sphere of radius r and a cylinder of radius R are

$M_{sph}(r) = \int_0^r dr\ 4\pi r^2 \rho(r) = 4\pi Cr$ and $M_{cyl}(R) = \int_0^R dR\ 2\pi R \sigma(R) = 2\pi^2 CR$.

For $r = R$, $M_{sph}/M_{cyl} = 2/\pi \sim 1$. This is only relevant for observations. Dynamically, even for disk geometry, the relevant quantity is $M_{sph}(r)$, as one can see by applying Gauss' law to appropriately shaped closed volumes, as emphasized earlier.

In general, (radio) line observations give data on the intensity (and possibly polarization) of the targetted emitting matter in very many velocity (as implied by shifted frequency) channels in each pixel ≡ restoring beam. These data can be viewed and plotted in many different ways. The two most popular presentations give (1) a (polarized or total) intensity contour map with colour-coded iso-velocity contours superposed (Figure 8; called *spider diagram*), and (2) a cut through such a map showing a plot of velocity vs distance along the cut (Figure 9)(called position-velocity or PV plot).

---

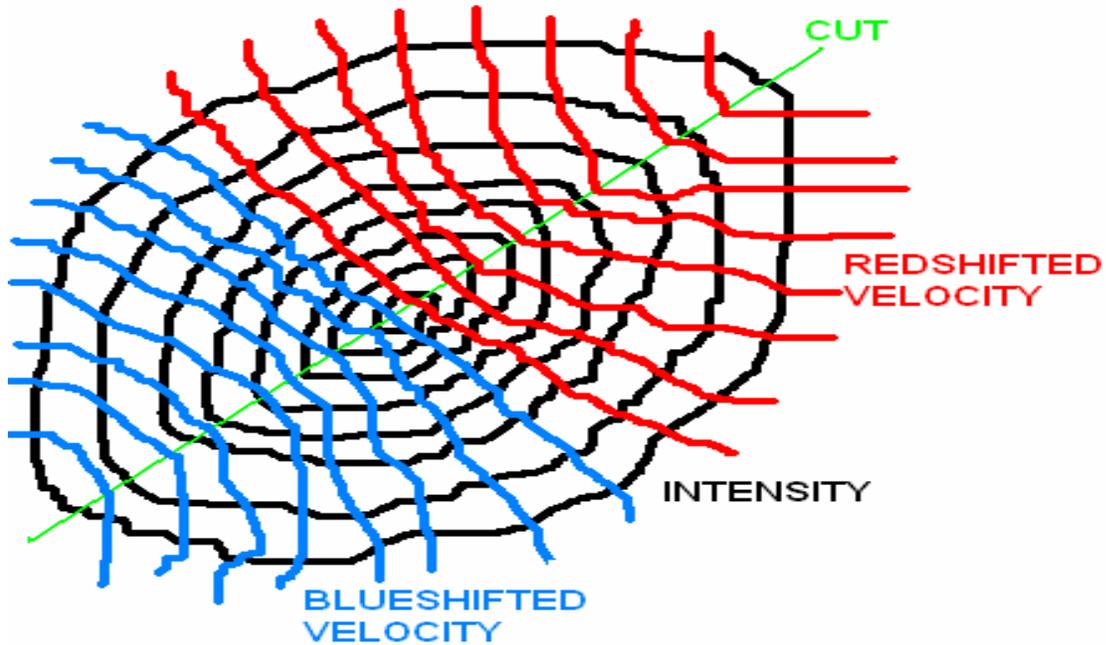

**Figure 8.** Schematic representation of iso-velocity contours superposed on intensity contour map (*spider diagram*)

___________________________________________________________________________

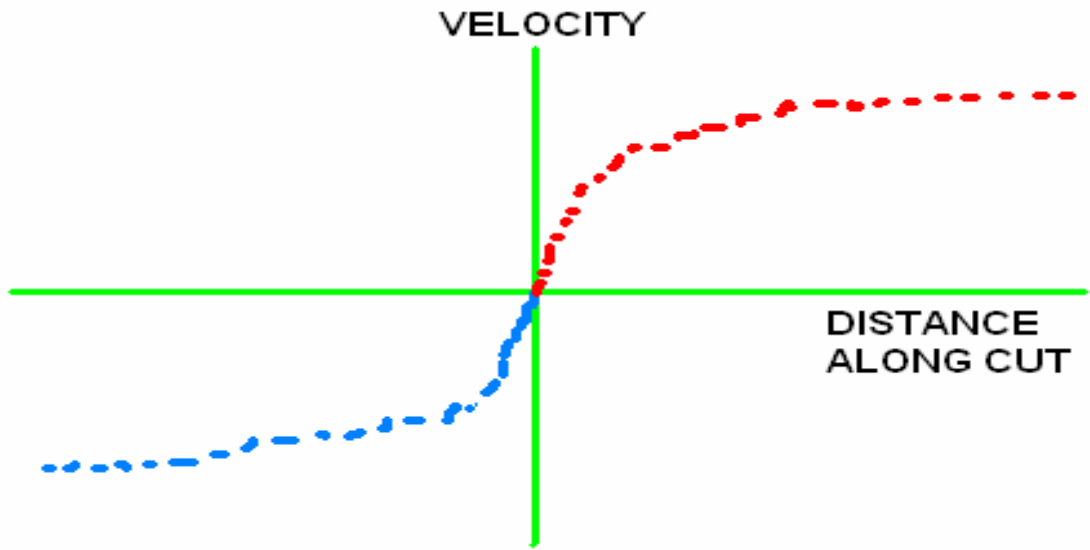

**Figure 9a.** Schematic galaxy rotation curve (called PV plot or position velocity plot), such as may be derived from a spider diagram like in Fig. 8 by integrating or taking a one-dimensional section.
___________________________________________________________________________

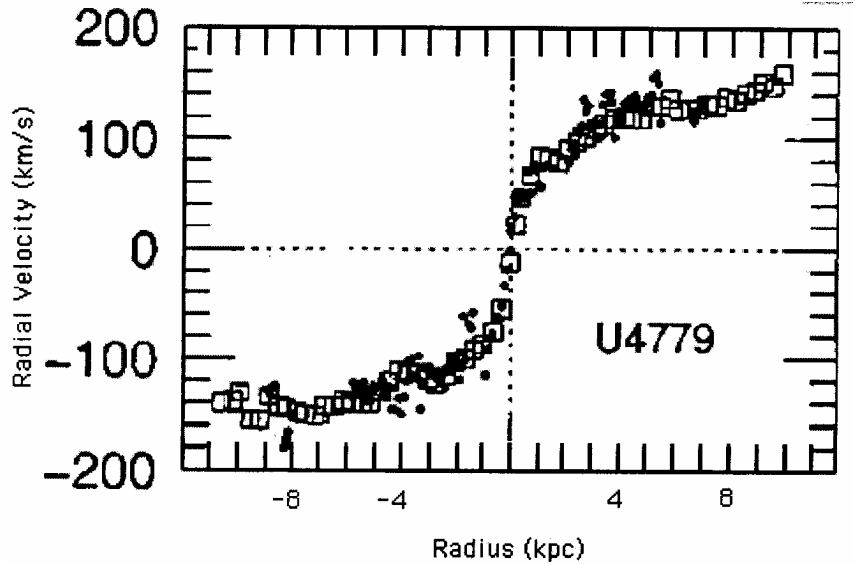

**Figure 9b.** An observed galaxy rotation curve (ref. **(26)**) for UGC4779 = NGC2742.

___

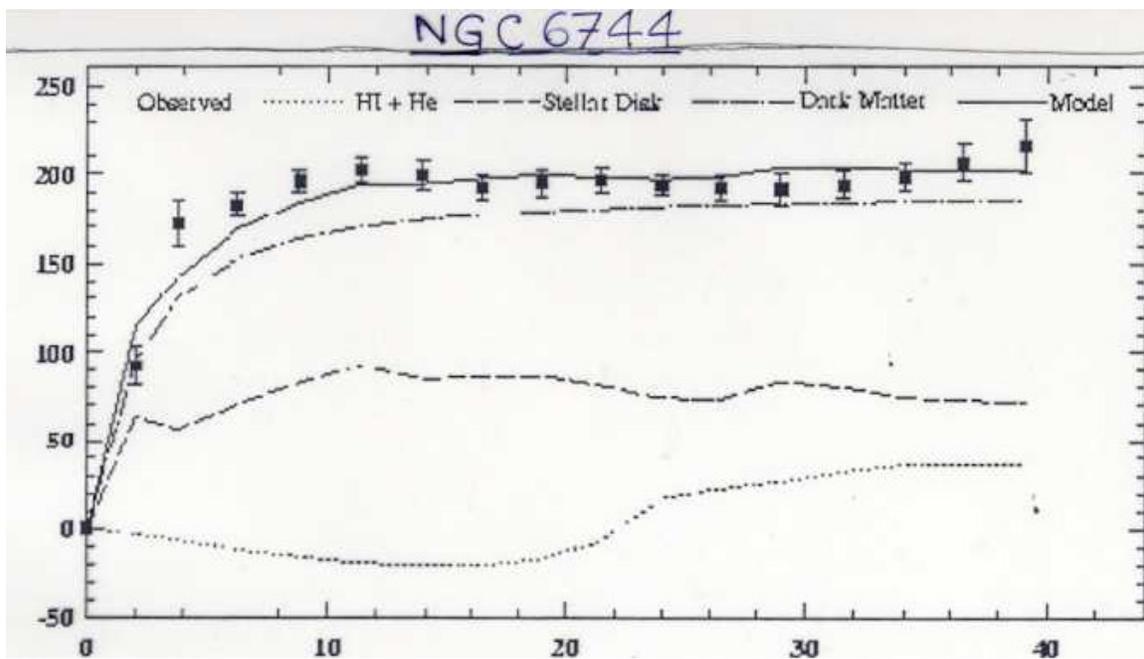

**Figure 10.** Rotation curve of NGC6744 to large galactocentric distance. (Velocity in km/s vs deprojected distance in kpc.) (From reference **(24)**.)

___

The velocity is always corrected for projection using an estimate of the inclination angle of the disk to the line of sight. There are other ways of getting the rotation curve, defined as such a position velocity plot[19]. A neutral hydrogen (i.e., $H^0$) study[24] of NGC 6744 out to 40 kpc is shown in Figure 10. UGC2885, NGC 5533 and NGC6674 rotation curves are measured to ~70 kpc[25].

In the radio band, rotation curves can be measured out to about 2 to 3 Holmberg radii[27], while in the optical band, it is possible to measure only to about 0.5 times Holmberg radius. The Holmberg radius corresponds to the isophote at a specific well-defined low surface brightness, about 1 to 2 % above the background sky brightness[17]. Other scales used to gauge the extent of disk galaxies are exponential disk scale length $r_d$ from I-band photometry and the radius $R_{opt}$ encompassing 83% of the total integrated light, fruitfully used by Catinella et al[28] who constructed template rotation curves by combining data on about 2200 disk galaxies, fitting for the amplitude $V_0$, exponential scale $r_{pe}$ of the inner region and the slope α of the outer part:

$$V_{pe}(r) = V_0[1 - \exp(-r/r_{pe})](1 + \alpha\, r/r_{pe}),$$

pe representing "polyex", the name of the model, and r & $r_{pe}$ expressed in units of $r_d$ or $R_{opt}$. Detailed mapping of radial distribution of visible and dark matter in disk galaxies requires use of luminosity profiles and extended $H^0$ rotation curves in addition to optical rotation curves. Martín[27] has collated $H^0$ maps from literature published between 1953 to 1995 into a uniform catalogue of about 1400 disk galaxies, and has analyzed some of the data in an attempt to derive features and relations common to most galaxies. It is sufficient to use angular distances (arcminutes or arcseconds) for the radial distances r, $r_{pe}$, $r_d$ and $R_{opt}$ for such compilations. A translation to physical units (kpc) is needed only for going to mass models. From a study of a homogeneous sample of about 1100 optical and radio rotation curves and relative surface photometry, Persic et al[29] found that a single property like total luminosity dictates the rotation velocity at any radius for any galaxy, revealing the existence of a universal rotation curve, which they derived, confirming their result from five years earlier, based on 58 rotation curves. However, Bosma[30] cautions that derivation of any universal rotation curve may not be warranted. See also recent evidence for two halo components in a galaxy, one of them retrograde[20]. I also mention Narayan & Jog[31] in this connection, who show how the convenience of presenting intensity versus radius as a log-linear plot led to a spurious observational cut-off to the luminous disk, promptly taken up by theorists to construct elaborate models! Different ways of presenting astrophysical data are essential for many purposes, as, for example, done by Kundt[32], who presents rotation curves in a novel way.

**(Newtonian) theory of (disk) galaxy rotation curves.** Binney & Tremaine[17] give a detailed treatment of observations, models and theory of galaxian dynamics within Newtonian gravitational and dynamical framework. Saslaw[33] has treated gravitational systems in general from a much broader perspective, and, in the process, given a concise summary of the essential Newtonian methodology. The reader can refer to these (and other possibly more recent) studies for details. Here only a few glimpses are given, hopefully enough to whet the appetite for more.

*Relation between Φ(r) and ρ(r) [or $V^2(r)$ and M(r)]*. The gravitational potential energy per unit mass is called *Newtonian gravitational potential* Φ. This is only a convenient mathematical quantity and has no physical existence in Newtonian theory, which is a simultaneous far-action theory rather than (special) relativistically propagating field theory like (classical) electromagnetism. (Cf. Graneau & Graneau[34] for an elaboration of this point. See also Banhatti & Banhatti[35].)

For a test particle in a circular orbit at radius r in a spherically symmetric mass distribution $\rho(r)$, the *circular speed* V(r) is found from

$$V^2(r) = rd\Phi/dr = rF = GM(r)/r = (4\pi G/r)\int_0^r dr'r'^2\rho(r'),$$

where F is the (radial) force / unit mass, and M(r) mass within a sphere of radius r. The *escape speed* $V_e = (2|\Phi(r)|)^{1/2}$, since the kinetic and potential energies are just balanced at this speed.

*Potential ↔ density pairs (& other quantities)*

*Spherically symmetric*. For illustration, some simple potentials are listed.

# Point mass M. $\Phi(r) = -GM/r$, $V(r) = (GM/r)^{1/2}$ & $V_e(r) = (2GM/r)^{1/2}$.

# Homogeneous sphere. $M(r) = (4/3)\pi r^3\rho$, with $\rho$ uniform (i.e., independent of r) & $V(r) = (4\pi G\rho/3)^{1/2}r$, rising linearly with radius. The orbital period is
$T = 2\pi r/V = (3\pi/G\rho)^{1/2}$, independent of radius r. A test mass released at r oscillates harmonically around r = 0, where it reaches after $t_{dyn} = T/4 = (3\pi/16G\rho)^{1/2}$, the *dynamical time* of a system of mean density $\rho$. For radial size a,

$$\Phi(r) = \begin{bmatrix} -2\pi G\rho(a^2 - r^2/3), & r \leq a \\ -4\pi G\rho a^3/3r, & r \geq a \end{bmatrix}.$$

# Other systems of interest are[17] isochrone potential, modified Hubble profile and power-law density.

* *Flattened systems*. Plummer-Kuzmin, Toomre's n and logarithmic. For details, see Binney & Tremaine[17].

*Poisson's equation for thin disks*. For an axisymmetric system with density $\rho(R,z)$,

$$\partial^2\Phi/\partial z^2 = 4\pi G\rho(R,z) + (1/R)(\partial/\partial R)(RF_R); F_R \equiv -\partial\Phi/\partial R \text{ being the radial force.}$$

Near z = 0, the first term on the RHS >> the second term, so that $\partial^2\Phi/\partial z^2 = 4\pi G\rho(R,z)$.

So Poisson's equation for a thin disk can be solved in two steps: (1) Using surface density (zero thickness), determine $\Phi(R,0)$. (2) At each radius R, solve this simplified Poisson's equation for the structure normal to the disk.

*Disk potentials*. By separating variables in cylindrical polar coordinates, surface density $\sigma(R)$ and potential $\Phi(R,z)$ are related by

$$\Phi(R,z) = -2\pi G\int_0^\infty dk\exp(-k|z|)J_0(kR)\{\int_0^\infty dR'R'\sigma(R')J_0(kR')\},$$

where $J_0$ is a Bessel function. Writing $S(k) = -2\pi G \int_0^\infty dR R \sigma(R) J_0(kR)$, the circular speed is given by

$V^2(R) = R(\partial \Phi/\partial R)_{z=0} = -R\int_0^\infty dk k S(k) J_1(kR)$, using $dJ_0(x)/dx = -J_1(x)$.

*Examples applying these formulae* (to get rotation curves from mass distributions)

#Rotation curve of Mestel's[36] disk: $\sigma(R) = \sigma_0 R_0/R$.

Calculation gives uniform, i.e., R-independent, circular speed: $V^2 = 2\pi G \sigma_0 R_0$.

Since $M(R) = 2\pi\int_0^R dR' R' \sigma(R')$, this can also be written $V^2 = GM(R)/R$, as for a spherical system, which is true only for Mestel's disk.

#Exponential disk: $\sigma(R) = \sigma_0 \exp(-R/R_d)$.

Calculation gives $V^2(R) = 4\pi G\sigma_0 R_d y^2 [I_0(y)K_0(y) - I_1(y)K_1(y)]$ where $y \equiv R/2R_d$, and $I_0$, $K_0$, $I_1$, $K_1$ are Bessel functions.

Deducing $\sigma(R)$ given $V(R)$ is formally possible, but involves differentiating the noisy observed function $V^2(R)$, which numerically worsens the error and is unstable.

**Fourier (numerical) method in cylindrical polar coordinates**[37]. With $u \propto \ell nR$, a rectangular grid in $(u,\varphi)$ plane, where $-\infty < u < \infty$ & $0 \leq \varphi$ (the azimuthal angle) $\leq 2\pi$, generates, in the $(R,\varphi)$ plane, cells that become smaller as $R \to 0$. This is well-suited to (numerical simulations of) centrally concentrated disks. A particularly efficient implementation is R. H. Miller's (1970s) code[38], using "leapfrog" numerical scheme, called Verlet method in molecular dynamics, extensively used at Turku (Finland) in mid-1980s and later[39] for various aspects of disk galaxies, including spiral arms, tidal interactions, Seyfert activity, etc. The N-body code uses 60000 particles in a smoothed potential. The particles are distributed on a 24 by 36 standard grid (part of which is shown in Figure11), determined by $r = L\exp(\lambda u)$; $\lambda = 2\pi/36$ and u ranging from 0 to 23.5, so that r ranges from L to 60.4L.

---

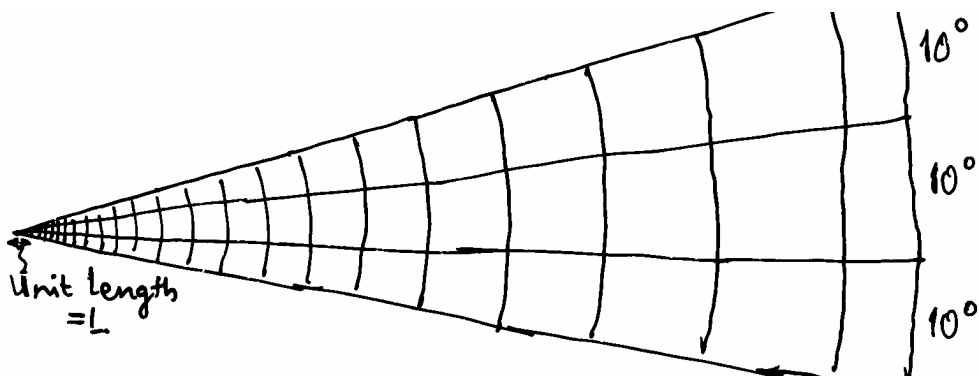

**Figure 11.** The two-dimensional polar coordinate grid used in R. H. Miller's numerical code[38].

-----------------------------------

The initial disk has

$$\sigma(R) = \begin{bmatrix} [V_0^2/\pi GR]\cos^{-1}(R/A) & \text{for } R \le A \\ 0 & \text{for } R > A \end{bmatrix}, \text{ where } A \equiv \text{disk radius.}$$

(see Grupen[1], eq. (13.7) on p.268, where one finds a concise summary of disk galaxy dynamics vis-a-vis motivation for dark matter.)

Thus the circular speed $V_0$ is, by integrating $\sigma(R)$ to get the total galaxy mass M, given by

$V_0^2 = \pi GM/2A$. The potential / unit mass is

$$\Phi(R) = \begin{bmatrix} V_0^2 \ln(R/2A) & \text{for } R \le A \\ -(2V_0^2/\pi)\sum_{\ell=0}^{\infty}[1\cdot 3\cdot 5\cdots(2\ell+1)/\{2^\ell \ell!(2\ell+1)^3\}](A/R)^{(2\ell+1)} & \text{for } R > A \end{bmatrix}.$$

The action of a spherical halo of density $\propto 1/r^2$ is the same as if it was projected on the disk plane, as can be verified by calculation that the surface density of the projected halo has the same form as assumed. In fact, this is the motivation for using this form. *The halo acts on the disk, but is not acted on, either by itself, or by the disk. This is somewhat puzzling, and may give spurious effects.*

**Disk-halo break-up & a new calculation scheme[7].** Poisson equation is numerically solved for an axially symmetric disk in cylindrical polar coordinates, using 250000 points = particles distributed along 500 rings of radii $\propto i^2$, where i is the ring number from centre outward, upto $R_g$, the radius of the finite disk. The force $\mathbf{F} \propto \int[\sigma(R)/R^3]\mathbf{R}dR$ acting on a given particle is discretized to

$\mathbf{F}_i = \sum_{j \ne i} G(m_i m_j / d_{ij}^3)\mathbf{d}_{ij}$, where $\mathbf{d}_{ij} = \mathbf{x}_j - \mathbf{x}_i$.

also $= m_i(v_i^2/d_i)(\mathbf{x}_i/d_i)$ to give rotation curve, $i = 0, 1, \ldots, n$.

$d_{ij}^2 = d_i^2 + d_j^2 - 2d_i d_j \cos(\theta_{ij})$, so that, with $F_{ij} = G(d_i - d_j\cos\theta_{ij})/d_{ij}^3$, the set of equations reduces to $\sum_{j \ne i} m_j F_{ij} = v_i^2/d_i$. This is a system of n linear equations, with n+1 unknowns $m_i$ (since we seek to invert $V^2 \to \sigma$). The additional equation needed is provided by the total mass $M_g$ of the galaxy: $M_g = \sum_i m_i$. Writing $\mu_i = m_i/M_g \equiv \omega m_i$, the n+1 equations for n+1 unknowns $\mu_i$, for each value of $\omega$, are (for $i = 1, \ldots, n$) ($i = 0$ gives $\mathbf{0} = \mathbf{0}$):

$\sum_{j=0}^{n} {}_{(j \ne i)} m_j F_{ij} = \omega v_i^2/d_i, \qquad \sum_{j=0}^{n} \mu_j = 1, \qquad$ with the constraint $\mu_i \ge 0$.

The constraint restricts $\omega$ between $\omega_{min}$ and $\omega_{max}$ which are close to each other (within $10^{-2}$ or less). The physical significance of the existence of the free parameter $\omega$ is that the rotation curve is known only upto $R_g$, the radius of the finite disk. The range allowed for ω corresponds to all possible extensions of the rotation curve beyond $R_g$. Since the range of ω is narrow, the mass of the galaxy from the known rotation curve is naturally found.

The method has been tested successfully for exponential disk, point mass and Mestel's disk. Figure 12 shows the surface density σ(R) for the Milky Way derived in this way from the rotation curve of Vallée[40]. $R_g$ = 14 kpc is used from Robin et al[41]. Comparison with other results and taking into account MACHO gravitational lensing candidates toward LMC, a halo of the same radius as the disk is needed.

However, as Méra et al[42] (& references therein) show in detail using star counts, microlensing observations and kinematics, the model eventually to be found consistent with all constraints is not yet determined, although both maximal halo type models with non-baryonic (i.e., exotic) dark matter and maximal disk type models with all matter baryonic are possible.

Gentile et al[43] use $H^0$ and Hα data on five spirals to decompose the rotation curve into stellar, gaseous and dark matter contributions, pointing toward halos with constant density cores.

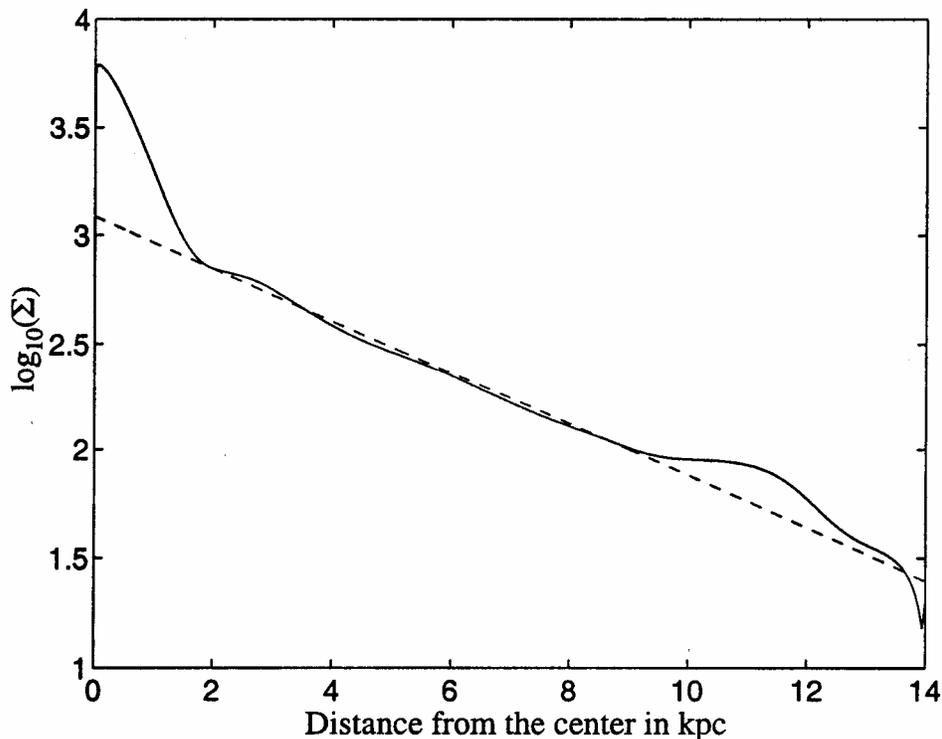

**Figure 12.** Surface mass density for Milky Way Galaxy compared with the model of Méra et al[1] (dotted). [The same figure is also available in Méra et al[42] as figure 2.]

**MoND and disk galaxy rotation curves**[44]. MoND was proposed in 1983 as a phenomenological model to fit gross features of rotation velocity data for spirals, especially Tully-Fisher relation M(mass) $\propto V^4$, V denoting the flat value of the outermost part of the rotation curve. The idea behind MoND was to seek a low field value $a_0$ ($\approx 10^{-8}$cm/sec$^2$, as it turns out), which modifies Newtonian gravitational acceleration (or field) from $g_N$ to $(g_N a_0)^{1/2}$ for low gravitational fields. As it happens, $a_0 \approx cH_0/6$, with c the speed of light and $H_0$ the Hubble constant. Detailed variation of spiral rotation curves fits very well with only one additional parameter giving the mass-to-light ratio across the whole spiral disk. Thus evidence for dark matter from spiral rotation curves can equally well be interpreted as evidence for MoND, which may at worst be taken to be a parametrization of rotation curve data, which a more fundamental model or theory should account for. The search is still on for such a physical basis (in the form of a field theory) for this acceleration (i.e., gravitational field)-based modification of Newtonian dynamics (and inertia). Recently[45], a unification of dark matter and dark energy into a dark fluid described by a tensor-vector-scalar (TeVeS) field theory has consonance with MoND. For details, see Sanders & McGaugh[42] and Zhao[45].

**General relativity vs. Newtonian gravity and dynamics.** Newtonian gravity & dynamics together form a simultaneous far-action theory. Newtonian gravitational potential (energy) is essentially a force field, giving the force / unit mass in space. This is true despite application of sophisticated mathematical techniques, as there is no (special) relativistic field propagation in Newtonian theory, which is not Lorentz-invariant, but Galilean, with absolute simultaneity, and space & time independent of each other. General relativity, on the other hand, is locally Lorentz-invariant, and is a genuine field theory with relativistic field propagation built into its structure.

*Galaxy rotation curves.* A galaxy is modeled as a stationary axially symmetric pressure-free fluid in general relativity. The rotation curve is derived by tracing paths of test particles, i.e., determining geodesics. The recent claim by Cooperstock & Tieu[46] that such a procedure leads to a good fit to observed galaxy rotation curves, without having to assume the existence of exotic dark matter, has been shown to be misguided by Vogt & Letelier[47], who infer that matter of negative energy density is implied in the disk. Korzyński[48] has shown that a singular disk is implied. Cross[49] found that correcting an internal inconsistency no longer leads to a flat rotation curve. Fuchs & Phelps[50] have pointed out the inadequacy of the model to reproduce local mass density and vertical density profile of the Milky Way. However, the idea of treating the nonlinear galacto-dynamical problem using general relativity rather than Newtonian gravity, due to the inherent nonlinearity of self-gravity and (general relativistic) dragging of inertial frames in the rotating model spacetimes, is well worth pursuing further. An example is Vogt & Letelier[51]. Another is Balasin & Grumiller[52], who point out that the Newtonian approximation breaks down globally, even though it is valid locally everywhere, confirming Fuchs & Phelps' failed attempt[50] to apply the model to the Milky Way Galaxy. Letelier[53] has summarized the position, also referring to an earlier relevant study by González & Letelier[54]. Finally, it is worth mentioning Gödel's study of rotating spacetimes, examined from a broader perspective by Yourgrau[55]. See also Banhatti[56].

**Summary / Conclusion.**

# Galaxy rotation curves are flat to the greatest extent they can be measured. Precise observations have delineated the undulations in this overall flat structure, especially for the Milky Way Galaxy. The low-field theory MoND phenomenologically fits details of disk rotation curves surprisingly well, with the same mass-to-light ratio across the whole disk as the only galaxian parameter, along with a constant low-field value, the constant acceleration of about $10^{-8}$cm/sec$^2$, intrinsic to MoND.

# There is a direct relation between rotation curves and mass models for galaxies. Dark matter is probably needed to fit the observations. In particular, a break-up of the mass distribution into a disk + a halo around the disk is probably needed, for both luminous and dark matter components.

# In addition to analytic approaches, numerical calculations / simulations in 2D polar coordinates are fruitful in relating mass models to observations.

# It is not clear if general relativistic treatment, normally needed for high gravitational fields, gives anything more than Newtonian dynamics, but is worth exploring further.

**Acknowledgments.** I thank Prof. Dr. Wim de Boer for inviting me to Karlsruhe, Germany for a visit and a seminar. University of Münster, Germany provided general facilities and use of library. Part of the work was done while visiting Institute of Mathematical Sciences (MatScience), Chennai, where I also gave a seminar on the topic. Department of Theoretical Physics, University of Madras also arranged my seminar on this topic. Nora Loiseau's comments at various stages improved clarity. Prof. Dr. Wolfgang Kundt, Bonn, Germany provided valuable comments. Prof. Peter Boschan at Münster encouraged me by discussing relevant matters and lending books when needed. Radha, my wife, has been a great friend all through, and provided physical, moral and spiritual support, and continues to do so. Moreover, I thank her for encouragement and critical comments on various versions of the article. I also thank the referees for their valuable suggestions.

**Notes added in proof :** **(1)** For an exposition on how a rotation curve is derived from the observations of a disk galaxy velocity field on the sky, deprojecting for disk inclination, see Teuben[57]. **(2)** Over the years there have been attempts to derive radial disk mass distributions from observed rotation curves. Two numerical schemes are described in the article. Another semi-analytic attempt is by Feng & Gallo[58], who also refer to a series of numerical calculations by K. F. Nicholson. All these independent studies do not need dark matter. **(3)** For an update on MoND after Sanders & McGaugh[44], see Bekenstein[59]. **(4)** Finally, a fundamental theory purporting to show that dark matter, dark energy and the accelerating universe are artifacts of our 3+1 dimensional perspective on a pre-geometric structure based on "process physics" has been developed by Cahill[60].

**References.**
**(1)** C. Grupen (2005) *Astroparticle Physics* (Springer) (Textbook).
**(2)** D. G. Banhatti (1993) *Current Science* **65** 827-35: Large-scale structure in the universe; (1994) *Physics Education* **11** 175-83: Cosmography; (1998) *Physics Education* **15** 273-82: Early universe & present large-scale structure – Part 1; (2000) *Physics Education* **17** 161-70: Early universe & present large-scale structure – Part 2.

**(3)** D. N. Spergel et al (2003) *Astrophysical Journal Supplement* **148** 175-94: First year WMAP observations; (2007) astro-ph/0603449v2: WMAP three-year observations.
**(4)** C. –A. Faucher-Giguere et al (2008) *Science* **319** 52-5: Numerical simulations unravel the cosmic web; J. F. Navarro et al (1997) *Astrophysical Journal* **490** 493-508: A universal density profile from hierarchical clustering; M. Kamionkowski & S. M. Koushiappas (2008) arXiv:0801.3269 astro-ph: Galactic substructure & direct detection of dark matter.
**(5)** B. D. Fields & S. Sarkar (2003) astro-ph/0406663v1: Big bang nucleosynthesis in S. Eidelman et al (2006) *Phys Lett* **B592** 1-1109 [http://pdg.lbl.gov] (Particle Data Group).
**(6)** R. J. Gaitskell (2004) *Ann Rev Nucl Part Sci* **54** 315-59: Direct detection of dark matter; L. M. Krauss (2007) astro-ph/0702051v1: Dark matter candidates – what's cold … & what's not; A. Bottino & N. Fornengo (1999) astro-ph/9904469: Dark matter & its particle candidates; G. Bertone et al (2005) *Physics Reports* **405** 279-390: Particle dark matter – evidence, candidates & constraints; J. Ellis (2000) *Physica Scipta* **T85** 221-30: Particle candidates for dark matter; M. Kamionkowski & S. M. Koushiappas (2008) arXiv:0801.3269 astro-ph: Galactic substructure & direct detection of dark matter.
**(7)** D. Méra, M. Mizony & J. –B. Baillon (1996/7) preprint (submitted to *Astronomy & Astrophysics* and *Monthly Notices of the Royal Astronomical Society*): Disk surface density profile of spiral galaxies and maximal disks.
**(8)** W. de Boer et al (2005) *Astronomy & Astrophysics* **444** 51-67: EGRET excess of diffuse galactic gamma rays as tracer of dark matter. [See also W. de Boer (December 2005) *CERN Courier* **45**(10) 17-19: Do gamma rays reveal our galaxy's dark matter? (Reading this was the beginning of my interest in this topic.)]
**(9)** Y. Xu et al (2006) *Science* **311** 54-57 + supporting online material: The distance to the Perseus spiral arm in the Milky Way.
**(10)** K. Hachisuka et al (2006) *Astrophysical Journal* **645** 337-344 (also astro-ph/0512226): Water maser motions in W3(OH) and a determination of its distance.
**(11)** A. M. Mel'nik (2006) *Astronomy Letters* **32**(1) 7-13: Outer pseudoring in the Galaxy [translated from the Russian (2006) *Pis'ma v Astronomicheskiĭ Zhurnal* **32**(1) 9-15].
**(12)** H. -W. Rix & D. Zaritsky (1995) *Astrophysical Journal* **447** 82-102: Nonaxisymetric structures in the stellar disks of galaxies.
**(13)** A. Dekel (2005) *Nature* **437** 707-10: Lost and found dark matter in elliptical galaxies.
**(14)** F. Aharonian et al (2006a) *Nature* **439** 695-698: Discovery of very-high-energy γ-rays from the Galactic Centre ridge; (2006b) *Physical Review Letters* **97** 221102.
**(15)** S. Ando (2007) *Jnl of Phys* **60** 247-50: Cosmic γ-ray background from dark matter annihilation.
**(16)** F. H. Shu (1982 or 1985) *The physical universe: an introduction to astronomy* University Science Books.
**(17)** J. Binney & S. Tremaine (1987) *Galactic dynamics* Princeton University Press.
**(18)** V. L. Fish et al (2003) *Astrophysical Journal* **587** 701-713: H I absorption toward ultracompact H II regions – distances and Galactic structure.
**(19)** Y. Sofue & V. Rubin (2001) *Annual Reviews of Astronomy & Astrophysics* **39** 137-74: Rotation curves of spiral galaxies.
**(20)** D. Carollo et al (2007) *Nature* **451** 1020-5 (+ erratum in 10Jan08 issue): Two stellar components in the halo of the Milky Way.

**(21)** J. A. R. Caldwell & J. P. Ostriker (1981) *Astrophysical Journal* **251** 61-87: The mass-distribution within our galaxy – a three component model.
**(22)** R. Cowsik et al (1996) *Physical Review Letters* **76**(21) 3886-3889: Dispersion velocity of Galactic dark matter particles.
**(23)** W. Dehnen & J. Binney (1998) *Monthly Notices of the Royal Astronomical Society* **294** 429-438: Mass models of the Milky Way.
**(24)** S. D. Ryder et al (1999) *Publications of the Astronomical Society of Australia* **16**(1) 84-88: An H I study of the NGC6744 system.
**(25)** R. H. Sanders (1996) *Astrophysical Journal* **473** 117-129: The published extended rotation curves of spiral galaxies: confrontation with modified dynamics.
**(26)** S. Courteau (1997) *Astronomical Journal* **114**(6) 2402-2427: Optical rotation curves & linewidths for Tully-Fisher applications.
**(27)** M. C. Martín (1998a) *Astronomy & Astrophysics Supplements* **131** 73-75: Catalogue of H I maps of galaxies. I. (Only available in electronic form at the CDS via anonymous ftp 130.79.128.5 or [http://cdsweb.u-strasbg.fr/Abstract.html]); M. C. Martín (1998b) *Astronomy & Astrophysics Supplements* **131** 77-87: Catalogue of H I maps of galaxies. II. Analysis of the data.
**(28)** B. Catinella et al (2006) *Astrophysical Journal* **640** 751-61 (also astro-ph/0512051): Template rotation curves for disk galaxies.
**(29)** M. Persic et al (1996) *Monthly Notices of the Royal Astronomical Society* **281** 27-47: The universal rotation curve of spiral galaxies – I. The dark matter connection.
**(30)** A. Bosma (1998) astro-ph/9812013v1: Dark matter in disc galaxies.
**(31)** C. A. Narayan & C. J. Jog (2003) *Astronomy & Astrophysics* **407** L59-62: The puzzle about the radial cut-off in galactic discs.
**(32)** Kundt, W. (2007) in 11th Marcel Grossmann Meeting WSPC Proceedings: The proposed black holes around us.
**(33)** W. C. Saslaw (1985) *Gravitational physics of stellar & galactic systems* (especially section 59) Cambridge University Press.
**(34)** P. Graneau & N. Graneau (1993) *Newton versus Einstein: How matter interacts with matter* Carlton/Affiliated East-West.
**(35)** R. D. Banhatti & D. G. Banhatti (1996) *Physics Education (India)* **12**(4) 377-79: Book review of **(34)** Graneau & Graneau (1993).
**(36)** L. Mestel (1963) *Monthly Notices of the Royal Astronomical Society* **126** 553-75: On the galactic law of rotation.
**(37)** G. Byrd et al (1986) *Monthly Notices of the Royal Astronomical Society* **220** 619-31: Dynamical friction on a satellite of a disc galaxy.
**(38)** R. H. Miller (1978) *Astrophysical Journal* **224** 32-38: On the stability of disklike galaxies in massive halos; (1978) *Astrophysical Journal* **223** 811-823: Numerical experiments on the stability of disklike galaxies; (1976) *Journal of Computational Physics* **21** 400-437: Validity of disk galaxy simulations; (1974) *Astrophysical Journal* **190** 539-542: Stability of a disk galaxy; (1971) *Journal of Computational Physics* **8**(3) 464-: Partial iterative refinements.
**(39)** M. J. Valtonen et al (1990) *Celestial Mechanics & Dynamical Astronomy* **48**(2) 95-113: Dynamical friction on a satellite of a disk galaxy: the circular orbit.
**(40)** J. P. Vallée (1994) *Astrophysical Journal* **437** 179-183: Galactic magnetism and the rotation curves of M31 and the Milky Way.

**(41)** A. C. Robin et al (1992) *Astronomy & Astrophysics* **265** 32-39: The radial structure of the galactic disc.
**(42)** D. Méra et al (1998) *Astronomy & Astrophysics* **330** 953-62: Towards a consistent model of the Galaxy. II. Derivation of the model.
**(43)** Gentile, G. et al (2004) *Monthly Notices of the Royal Astronomical Society* **351**(3) 913-22: The cored distribution of dark matter in spiral galaxies.
**(44)** R. H. Sanders & S. S. McGaugh (2002) *Ann Rev Astr Astrophys* **40** 263-317: Modified Newtonian dynamics as an alternative to dark matter.
**(45)** H. S. Zhao (2007) arXiv:0710.3616v2 astro-ph: Coincidences of dark energy with dark matter – clues for a simple alternative?
**(46)** F. I. Cooperstock & S. Tieu (2005a) astro-ph/0507619: General relativity resolves galactic rotation without exotic dark matter. See also (2005b) astro-ph/0512048: Perspectives on Galactic dynamics via general relativity; (2006) astro-ph/0610370: Galactic dynamics via general relativity - a compilation and new developments.
**(47)** D. Vogt & P. S. Letelier (2005b) astro-ph/0510750: Presence of exotic matter in the Cooperstock & Tieu galaxy model. Also (2006) astro-ph/0611428: Exact general relativistic rotating disks immersed in rotating dust generated from van Stockum solution.
**(48)** M. Korzyński (2005) astro-ph/0508377: Singular disk of matter in the Cooperstock-Tieu galaxy model.
**(49)** D. J. Cross (2006) astro-ph/0601191: Comments on the Cooperstock-Tieu galaxy model.
**(50)** B. Fuchs & S. Phelps (2006) *New Astronomy* **11**(6) 608-610: Comment on **(46)** Cooperstock & Tieu (2005a).
**(51)** D. Vogt & P. S. Letelier (2005a) *Monthly Notices of the Royal Astronomical Society* **363** 268-84: Relativistic models of galaxies.
**(52)** H. Balasin & D. Grumiller (2006) astro-ph/0602519: Significant reduction of galactic dark matter by general relativity.
**(53)** P. S. Letelier (2006) in *IAU Symposium 238* 401- : Rotation curves, dark matter and general relativity.
**(54)** G. A. González & P. S. Letelier (2000) *Physical Review* **D62** 064025: Rotating relativistic thin disks.
**(55)** P. Yourgrau (2005) *A World without Time: the forgotten legacy of Gödel & Einstein*, Basic Books.
**(56)** D. G. Banhatti (2006) *Current Science* **90**(12) 1694: Book review of **(55)** Yourgrau (2005).
-----------------------------------------------------------------------
**(57)** P. J. Teuben {arXiv:astro-ph/020447v1 27 Apr 2002} : Velocity fields of disk galaxies (in *Disks of Galaxies : Kinematics, Dynamics & Perturbations*, ASP Conference Series, edited by E. Athanassoula & A. Bosma).
**(58)** J. Q. Feng & C. F. Gallo {arXiv:0803.0556v1 [astro-ph] 4 Mar 2008} : Galactic rotation described with thin-disk gravitational model.
**(59)** J. D. Bekenstein {arXiv:astro-ph/0701848v2 27 Mar 2007} : The modified Newtonian dynamics – MOND and its implications for new physics.
**(60)** R. T. Cahill {arXiv:0709.2909v1 [physics.gen-ph] 18 Sep 2007} : A quantum cosmology: no dark matter, dark energy nor accelerating universe.
=======================================